\documentclass[conference]{IEEEtran}
\IEEEoverridecommandlockouts

\usepackage{amssymb}
\usepackage[cmex10]{amsmath}
\usepackage{stfloats}
\usepackage{graphicx}
\usepackage{subfigure}
\usepackage{tabularx}
\usepackage{epsfig,epsf,color,balance,cite}
\usepackage{verbatim}
\usepackage{url}
\usepackage{bm}
\usepackage{booktabs}
\usepackage[top=0.75in, bottom=1.03in, left=0.64in, right=0.64in]{geometry}

\usepackage{algorithm}
\usepackage{algorithmic}

\usepackage{color}
\definecolor{myc1}{rgb}{0,0,0}

\begin{document}

\title{Fluid Antenna Relay Assisted Communication Systems Through Antenna Location Optimization}

\author{
\IEEEauthorblockN{
Ruopeng Xu$\IEEEauthorrefmark{1}$$\IEEEauthorrefmark{2}$,
Yixuan Chen$\IEEEauthorrefmark{1}$$\IEEEauthorrefmark{2}$,
Jiawen Kang$\IEEEauthorrefmark{3}$,
Minrui Xu$\IEEEauthorrefmark{4}$,
Zhaohui Yang$\IEEEauthorrefmark{1}$$\IEEEauthorrefmark{2}$,
Chongwen Huang$\IEEEauthorrefmark{1}$$\IEEEauthorrefmark{2}$,
Dusit Niyato$\IEEEauthorrefmark{4}$}
	\IEEEauthorblockA{
			$\IEEEauthorrefmark{1}$College of Information Science and Electronic Engineering, Zhejiang University, Hangzhou, China\\
			$\IEEEauthorrefmark{2}$Zhejiang Provincial Key Laboratory of Info. Proc., Commun. \& Netw. (IPCAN), Hangzhou, China\\
   $\IEEEauthorrefmark{3}$School of Automation, Guangdong University of Technology, China\\
   $\IEEEauthorrefmark{4}$School of Computer Science and Engineering, Nanyang Technological University,\\ Singapore 639798, Singapore\\
          	E-mails:
ruopengxu@zju.edu.cn,
chenyixuan@zju.edu.cn,
kavinkang@gdut.edu.cn,
minrui001@e.ntu.edu.sg, \\
 yang\_zhaohui@zju.edu.cn,
 chongwenhuang@zju.edu.cn,
 dniyato@ntu.edu.sg
		}
\thanks{The work was supported by the China National Key R\&D Program under Grant 2021YFA1000500, 2023YFB2904804 and 2023YFB2904800, National Natural Science Foundation of China under Grant 62331023, 62101492, 62394292, 62394290 and U20A20158, Zhejiang Provincial Natural Science Foundation of China under Grant LR22F010002, Zhejiang Provincial Science and Technology Plan Project under Grant 2024C01033, and Zhejiang University Global Partnership Fund.}
\vspace{-2em}
}
\maketitle

\begin{abstract}
 In this paper, we investigate the problem of resource allocation for fluid antenna relay (FAR) system with antenna location optimization.  In the considered model, each user transmits information to a base station (BS) with help of FAR.  The antenna location of the FAR is flexible and can be adapted to dynamic location distribution of the users.  We formulate a sum rate maximization problem  through jointly optimizing the antenna location and bandwidth allocation with meeting the minimum rate requirements, total bandwidth budget, and feasible antenna region constraints. To solve this problem, we obtain the optimal bandwidth in closed form.  Based on the optimal bandwidth, the original problem is reduced to the antenna location optimization problem and an alternating algorithm is proposed. Simulation results verify the effectiveness of the proposed algorithm and the sum rate can be increased by up to 125\% compared to the conventional schemes. 
 \end{abstract}

\IEEEpeerreviewmaketitle

\section{Introduction}
As an emerging technology, fluid antenna system (FAS) has recently gained a lot of attention from both industry and academia \cite{wong2020fluid,wong2022bruce,wong2020fluid2}. 
In FAS, the physical location of each antenna can be flexibly adjusted to one position in the discrete or continuous space, which can provide the highest channel gain.
Through proper antenna location optimization, only one antenna in the FAS can achieve the similar performance of the traditional multiple-antenna system.
Due to the above distinction, FAS presents a new direction to complement or even surpass multiple-input multiple-output (MIMO) \cite{chai2022port,konca2015frequency,waqar2023deep,wong2022closed,skouroumounis2022fluid,kelley2013frequency}. Specifically, the recent work in FAS  has considered the possibility of switchable antennas at the user equipment location to enhance the performance of wireless communication systems \cite{zheng2023flexible,wong2023fluid,wong2023fluid2,abu2021liquid,mukherjee2022level,borda2017low,paracha2019liquid,vega2023simple}.
The work about FAS can be classified into two aspects, performance analysis and system optimization.

As for performance analysis, 
the authors in \cite{wong2020fluid} first studied this type of FAS. Under the idealized model, they derived the precise and approximate expressions of the outage probability and the upper limit of the outage probability, and concluded that the size and number of ports of the fluid antenna are critical to single-user point-to-point systems influencing on the outage probability. In an ideal case, an infinitely small interruption probability can be achieved when the number of ports approaches infinity. Because the fluid antenna has a large number of ports and is distributed in a limited linear space, there is a strong spatial correlation between each port.
To handle this correlation issue, the work in
\cite{10103838} proposed a joint correlation channel model in order to more accurately consider the spatial correlation between the ports of the FAS in performance analysis, and the model of the fluid antenna has been continuously improved.
Recently, channel estimation of FAS and port selection in simple cases have also been studied in \cite{skouroumounis2022fluid}. Fluid antennas and MIMO can be combined to further improve system performance, giving rise to the concept of MIMO-FAS. Fluid antennas can provide additional degrees of freedom (DoF) for MIMO to achieve more spatial diversity.
Furthermore, a low-complexity channel reconstruction method for a FAS system was investigated in 
\cite{10375559}, where the base station (BS) utilizes
 a fixed multi-antenna uniform linear array (ULA) while each mobile user is equipped with a linear FAS.

As for  system optimization, 
the information theory performance of MIMO-FAS was studied in \cite{new2023information}, and the results show that MIMO-FAS is better than traditional MIMO. FAS is also effective for multi-user communication.
The  fluid antenna multiple access (FAMA) was proposed in \cite{wong2021fluid2} to serve multiple users in the domain of the fluid antenna.
By changing the port, the fluid antenna will be able to obtain the ups and downs of its interference signal in the spatial domain.
To show the effectiveness of FAS with multiple access schemes,
the sum rate of both 
orthogonal multiple access (OMA) and non-orthogonal multiple access (NOMA) networks were investigated in \cite{10318134} for FAS.
To enhance the security of the system, the secrecy rate maximization problem was investigated in 
\cite{10092780} for both perfect and imperfect CSI scenarios.
However, the above works \cite{new2023information,wong2021fluid2,10318134,10092780} all ignored the potential of using fluid antenna to serve as relay, which can greatly improve the system performance in particular for the cases with multiple obstacles.

To the authors' best knowledge, this is the first work to investigate
fluid antenna relay (FAR) assisted communication system, where fluid antenna is served as a relay. 
The main contributions of this paper are listed as follows:
\begin{itemize}
    \item We investigate an uplink FAR assisted communication system, where multiple users communication with the BS. In the considered model, the FAR with flexible antenna location is deployed inside the obstacle between the user and the BS. We then formulate an  optimization problem   to maximize the sum rate of FAR assisted system through jointly optimizing antenna location and bandwidth allocation.
    \item To solve this sum rate maximization problem, the optimal bandwidth is obtained in closed form as a function of the antenna location. Through substituting the optimal solution of the bandwidth allocation, we transform the  original joint antenna location and bandwidth allocation problem  to the antenna location problem. To solve the antenna location problem, an alternating algorithm with low complexity is proposed through iteratively solving a set of convex subproblems. 
    \item Simulation results show the superiority of the proposed algorithm. Compared to the conventional schemes, it is observed that the proposed scheme can achieve by up to 125\% gains in sum rate. 
\end{itemize}

The remainder of this paper is organized as follows. The system model and problem formulation are shown in Section \ref{section2}. The algorithm design is presented in Section \ref{section3}. Simulation results are analyzed in Section \ref{section4}, while conclusions are provided in Section \ref{section5}.
\addtolength{\topmargin}{-0.01in}
\section{System Model and Problem Formulation} \label{section2}
\begin{figure}[t]
    \centering
    \includegraphics[width=0.75\linewidth]{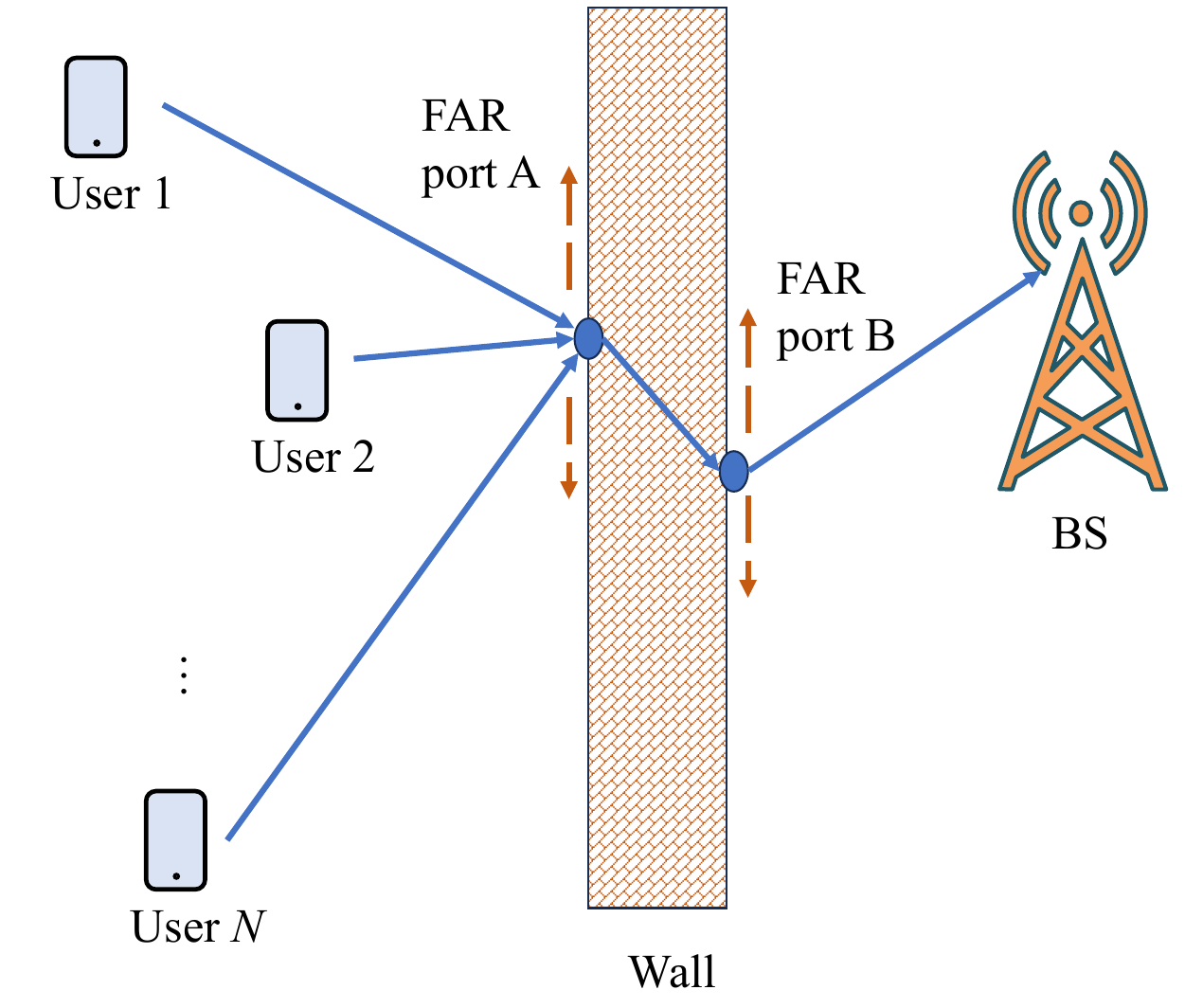}
    \caption{An uplink communication system with one BS, one FAR, and multiple users.}
    \vspace{-1em}
    \label{fig1}
\end{figure}

Consider an uplink system with one BS, one FAR and $N$ users, as shown in Fig.~\ref{fig1}. 
The set of users is denoted by the set $\mathcal N=\{1, \cdots, N\}$. 
Due to blockage such as walls in the building, there does not exist a line of sight (LoS) link between each user and the BS. 
To handle this issue, the FAR is deployed at the surface of the blocking wall, which includes two parts, i.e., port A and port B, as illustrated in Fig.~\ref{fig1}.
With the help of FAR, the communication link between users and the BS can be greatly improved, through establishing the virtual LoS user-port A-port B-BS link. 

\subsection{Communication Model}
Consider the three-dimensional location of the users, FAR, and BS. 
Since the user usually has low height, and the location of user $n$ is denoted by $(u_{n1}, u_{n2}, 0)$.
The location of the BS is denoted by $(s_1, s_2, H)$.
For the FAR with width $X$, the 
horizontal locations of both ports A and B are assumed to be fixed.
Thus, the locations of FAR ports A and B can be respectively denoted by $(0, y_1, z_1)$ and $(X, y_2, z_2)$, where $y_1, z_1, y_2$, and $z_2$ are flexible due to the design property of fluid antenna. 

Due to the fact that links users-port A, port A-port B, port B-BS are all LoS, the effective channel gain  from user $n$ to the BS through the help of FAR with  ports A and  B can be  given by 
\begin{align}\label{sys2eq1}
    &h_{n}( y_1, z_1,y_2, z_2)
    \nonumber\\
    =& \rho_0 \left( d_{n1}( y_1, z_1)+\frac{d_2( y_1, z_1, y_2, z_2)}{A}+d_3(y_2, z_2)\right)^{-\alpha},
\end{align}
where 
\begin{align}\label{sys2eq2}
d_{n1}( y_1, z_1)=\sqrt{u_{n1}^2+(y_1-u_{n2})^2+z_1^2},
\end{align}
\begin{align}\label{sys2eq3}
d_{2}( y_1, z_1,y_2, z_2)=\sqrt{X^2+(y_1-y_2)^2+(z_1-z_2)^2},
\end{align}
\begin{align}\label{sys2eq5}
d_{3}( y_2, z_2)=\sqrt{(X-s_1)^2+(y_2-s_2)^2+(z_2-H)^2},
\end{align}
and the constant parameter $A>1$ is introduced due to the difference of Electromagnetic waves propagation from solid media.
In \eqref{sys2eq1}, $d_{n1}( y_1, z_1)$ stands for the distance between user $n$ and FAR port A, $d_{2}( y_1, z_1,y_2, z_2)$ stands for the distance between FAR port A and AR port B,
$d_{3}( y_2, z_2)$ stands for the distance between FAR port B and the BS, $\rho_0$ is the reference channel gain at the distance of 1 meter, and $\alpha$ is the path loss parameter. 

Based on the channel gain given in \eqref{sys2eq1}, the achievable rate between user $n$ and the BS is 
\begin{align}\label{sys2eq5_2}
    r_n=b_n \log_2\left(1+ \frac{p_n h_{n}( y_1, z_1,y_2, z_2)}{\sigma^2}
    \right),
\end{align}
where $b_n$ is the bandwidth allocated to user $n$,
$p_n$ is the transmit power of user $n$,
and $\sigma^2$ is the Gaussian noise power.
\subsection{Problem Formulation}
Given the considered FAR model, our aim is to maximize the sum rate of the system considering the minimum rate requirements, bandwidth allocation, and antenna location constraints.
Mathematically, the sum rate maximization problem can be formulated as 
\begin{subequations}\label{sys2max0} 
	\begin{align}
		\mathop{\max}_{ \boldsymbol b,  y_1, z_1,y_2, z_2} \;&  
		 \sum_{n=1}^N b_n \log_2\left(1+ \frac{p_n h_{n}( y_1, z_1,y_2, z_2)}{\sigma^2} \right)\\
		\textrm{s.t.} \quad\:
		& b_n \log_2\left(1+ \frac{p_n h_{n}( y_1, z_1,y_2, z_2)}{\sigma^2} \right)\geq R_n, \quad \forall n,
		\\& \sum_{n=1}^N b_n \leq B, 
		\\& Y_{\min}\leq y_1,y_2\leq Y_{\max}, 
		\\&  Z_{\min}\leq z_1,z_2\leq Z_{\max}, 
    \\& b_n\geq0, \quad \forall n,
	\end{align}
\end{subequations}
where $\boldsymbol b=[b_{1},\cdots,b_{N}]^T$, $R_n$ is the minimum rate requirement for user $n$, $B$ is the total bandwidth of the system, $[Y_{\min}, Y_{\max}]$ is the feasible location region for $y_1$ and $y_2$, and $[Z_{\min}, Z_{\max}]$ is the feasible location region for $z_1$ and $z_2$.
Constraint (\ref{sys2max0}b) ensures the minimum rate requirement for each user. 
The total bandwidth of the system is limited as shown in (\ref{sys2max0}c).
Due to limited area of the building surface, the feasible location region of the antenna location for ports A and B can be presented by constraints  (\ref{sys2max0}d) and  (\ref{sys2max0}e).

Problem \eqref{sys2max0} is hard to solve due to the following two difficulties.
The first difficulty lies in that both bandwidth variable $\boldsymbol b$ and location optimization variables $(y_1, z_1,y_2, z_2)$ are coupled in the rate expression as shown in the objective function (\ref{sys2max0}a) and rate constraint (\ref{sys2max0}b).
The second difficulty is the complicated expression of the objective function with respect to the location optimization variables $(y_1, z_1,y_2, z_2)$ according to \eqref{sys2eq1}.

\section{Algorithm Design}\label{section3}
In this section, the optimal bandwidth allocation is first obtained. Then, through substituting the optimal expression of the bandwidth allocation into the original optimization problem, the joint bandwidth allocation and antenna location optimization problem is equivalent to the optimization problem only related to the antenna location optimization. For the antenna location optimization, the optimal location for FAR port B is obtained in closed form and an iterative algorithm is proposed to solve the antenna location for FAR port A. 

\subsection{Bandwidth Optimization}
With given antenna location optimization variables $(y_1, z_1,y_2, z_2)$ in problem \eqref{sys2max0}, the bandwidth allocation problem can be formulated as
\begin{subequations}\label{alo31max0} 
	\begin{align}
		\mathop{\max}_{ \boldsymbol b} \;&  
		 \sum_{n=1}^N b_n \log_2\left(1+ \frac{p_n h_{n}( y_1, z_1,y_2, z_2)}{\sigma^2} \right)\\
		\textrm{s.t.} \quad\:
		& b_n \log_2\left(1+ \frac{p_n h_{n}( y_1, z_1,y_2, z_2)}{\sigma^2} \right)\geq R_n, \quad \forall n,
  \\& \sum_{n=1}^N b_n \leq B, 
  \\& b_n\geq0, \quad \forall n,
	\end{align}
\end{subequations}
which is a linear optimization problem. 
Considering the property of linear optimization problem, it is clear that the additional bandwidth should be allocated to the user with the highest channel gain. 
Denote the user with the highest channel gain by
\begin{align}\label{alo31eq1}
    k=\arg\max_{n=1,\cdots,N} h_{n}( y_1, z_1,y_2, z_2).
\end{align}
Since the objective function is an increasing function of bandwidth, for the optimal solution of problem \eqref{alo31max0}, constraint (\ref{alo31max0}b) always holds with equality, i.e.,
\begin{align}\label{alo31eq1_2}
    \sum_{n=1}^N b_n^* = B.
\end{align}
Further considering rate allocation constraint (\ref{alo31max0}a), the optimal bandwidth can be given by 
\begin{align}\label{alo31eq2}
    b_n^*=\frac{R_n}{\log_2\left(1+ \frac{p_n h_{n}( y_1, z_1,y_2, z_2)}{\sigma^2} \right)}, \forall n \in \mathcal N, n \neq k.
\end{align}
Combining \eqref{alo31eq1_2} and \eqref{alo31eq2}, we have 
\begin{align}\label{alo31eq3}
    b_k^*=B-\sum_{n \in \mathcal N, n \neq k}\frac{R_n}{\log_2\left(1+ \frac{p_n h_{n}( y_1, z_1,y_2, z_2)}{\sigma^2} \right)}.
\end{align}

\subsection{Antenna Location Optimization}

Substituting the optimal bandwidth \eqref{alo31eq2} and \eqref{alo31eq3} into the original problem \eqref{sys2max0}, the joint bandwidth allocation and antenna location optimization problem \eqref{sys2max0} is equivalent to the optimization problem
\begin{subequations}\label{alo3max1} 
	\begin{align}
		\mathop{\max}_{ y_1, z_1,y_2, z_2} \;&  
		 \left(B-\sum_{n \in \mathcal N, n \neq k}\frac{R_n}{\log_2\left(1+ \frac{p_n h_{n}( y_1, z_1,y_2, z_2)}{\sigma^2} \right)}\right)
   \nonumber \\
   &\log_2\left(1+ \frac{p_k h_{k}( y_1, z_1,y_2, z_2)}{\sigma^2} \right)
   \\
		\textrm{s.t.} \quad\:
		& \left(B-\sum_{n \in \mathcal N, n \neq k}\frac{R_n}{\log_2\left(1+ \frac{p_n h_{n}( y_1, z_1,y_2, z_2)}{\sigma^2} \right)}\right)
   \nonumber \\
   &\log_2\left(1+ \frac{p_k h_{k}( y_1, z_1,y_2, z_2)}{\sigma^2} \right) \geq R_k, 
		\\& B-\sum_{n \in \mathcal N, n \neq k}\frac{R_n}{\log_2\left(1+ \frac{p_n h_{n}( y_1, z_1,y_2, z_2)}{\sigma^2}\right)}\geq 0 , 
  \\& k=\arg\max_{n=1,\cdots,N} h_{n}( y_1, z_1,y_2, z_2),
		\\& Y_{\min}\leq y_1,y_2\leq Y_{\max}, 
		\\&  Z_{\min}\leq z_1,z_2\leq Z_{\max}, 
    \\& b_n\geq0, \quad \forall n.
	\end{align}
\end{subequations}
Due to the fact that the objective function is to maximize the rate for user $k$ with the largest channel gain, minimum rate constraint (\ref{alo3max1}b) can be omitted.
The user $k$ with the largest channel gain should satisfy constraint (\ref{alo3max1}d).
Constraint (\ref{alo3max1}d) is complicated since the user index $k$ is a function of the antenna  location variables. 
To handle this issue, one can exhaustively search the user index $k$ with $N$ times as given in the following Algorithm 1 and then constraint (\ref{alo3max1}d) can be disregarded. 
As a result, the antenna location optimization problem can be formulated as:
\begin{subequations}\label{alo3max2} 
	\begin{align}
		\mathop{\max}_{ y_1, z_1,y_2, z_2} \;&  
		 \left(B-\sum_{n \in \mathcal N, n \neq k}\frac{R_n}{\log_2\left(1+ \frac{p_n h_{n}( y_1, z_1,y_2, z_2)}{\sigma^2} \right)}\right)
   \nonumber \\
   &\log_2\left(1+ \frac{p_k h_{k}( y_1, z_1,y_2, z_2)}{\sigma^2} \right)
   \\
		\textrm{s.t.} \quad\:
		&  B-\sum_{n \in \mathcal N, n \neq k}\frac{R_n}{\log_2\left(1+ \frac{p_n h_{n}( y_1, z_1,y_2, z_2)}{\sigma^2}\right)}\geq 0 , 
		\\& Y_{\min}\leq y_1,y_2\leq Y_{\max}, 
		\\&  Z_{\min}\leq z_1,z_2\leq Z_{\max}.
	\end{align}
\end{subequations}

The core of solving problem \eqref{alo3max2} lies in the channel gain expression $h_{n}( y_1, z_1,y_2, z_2)$ as shown in \eqref{sys2eq1}. 
According to \eqref{sys2eq1}, it is observed that antenna location variable of port B $(y_2, z_2)$ is only related to the BS and the location of port A $(y_1, z_1)$.
Considering the fact that the  Electromagnetic waves propagates faster in  solid medium than the gas medium, thus resulting that the value of $h_{n}( y_1, z_1,y_2, z_2)$ is mostly determined by distances $d_{n1}( y_1, z_1)$ and $d_{3}( y_2, z_2)$.
Based on this observation the antenna location optimization problem  \eqref{alo3max2} can be transformed into two subproblems,  antenna location optimization subproblem for port B
\begin{subequations}\label{alo3max31} 
	\begin{align}
		\mathop{\max}_{y_2, z_2} \;&  
		 \left(B-\sum_{n \in \mathcal N, n \neq k}\frac{R_n}{\log_2\left(1+ \frac{p_n h_{n}( y_1, z_1,y_2, z_2)}{\sigma^2} \right)}\right)
   \nonumber \\
   &\log_2\left(1+ \frac{p_k h_{k}( y_1, z_1,y_2, z_2)}{\sigma^2} \right)
   \\
		\textrm{s.t.} \quad\:
		&  B-\sum_{n \in \mathcal N, n \neq k}\frac{R_n}{\log_2\left(1+ \frac{p_n h_{n}( y_1, z_1,y_2, z_2)}{\sigma^2}\right)}\geq 0 , 
		\\& Y_{\min}\leq y_2\leq Y_{\max}, 
		\\&  Z_{\min}\leq z_2\leq Z_{\max}, 
	\end{align}
\end{subequations}
and antenna location optimization subproblem for port A
\begin{subequations}\label{alo3max32} 
	\begin{align}
		\mathop{\max}_{ y_1, z_1} \;&  
		 \left(B-\sum_{n \in \mathcal N, n \neq k}\frac{R_n}{\log_2\left(1+ \frac{p_n h_{n}( y_1, z_1,y_2, z_2)}{\sigma^2} \right)}\right)
   \nonumber \\
   &\log_2\left(1+ \frac{p_k h_{k}( y_1, z_1,y_2, z_2)}{\sigma^2} \right)
   \\
		\textrm{s.t.} \quad\:
		&  B-\sum_{n \in \mathcal N, n \neq k}\frac{R_n}{\log_2\left(1+ \frac{p_n h_{n}( y_1, z_1,y_2, z_2)}{\sigma^2}\right)}\geq 0 , 
		\\& Y_{\min}\leq y_1\leq Y_{\max}, 
		\\&  Z_{\min}\leq z_1\leq Z_{\max}.
	\end{align}
\end{subequations}
\addtolength{\topmargin}{-0.01in}
\subsubsection{Antenna Location Optimization for Port B}
\begin{figure}[t]
    \centering
    \includegraphics[width=\linewidth]{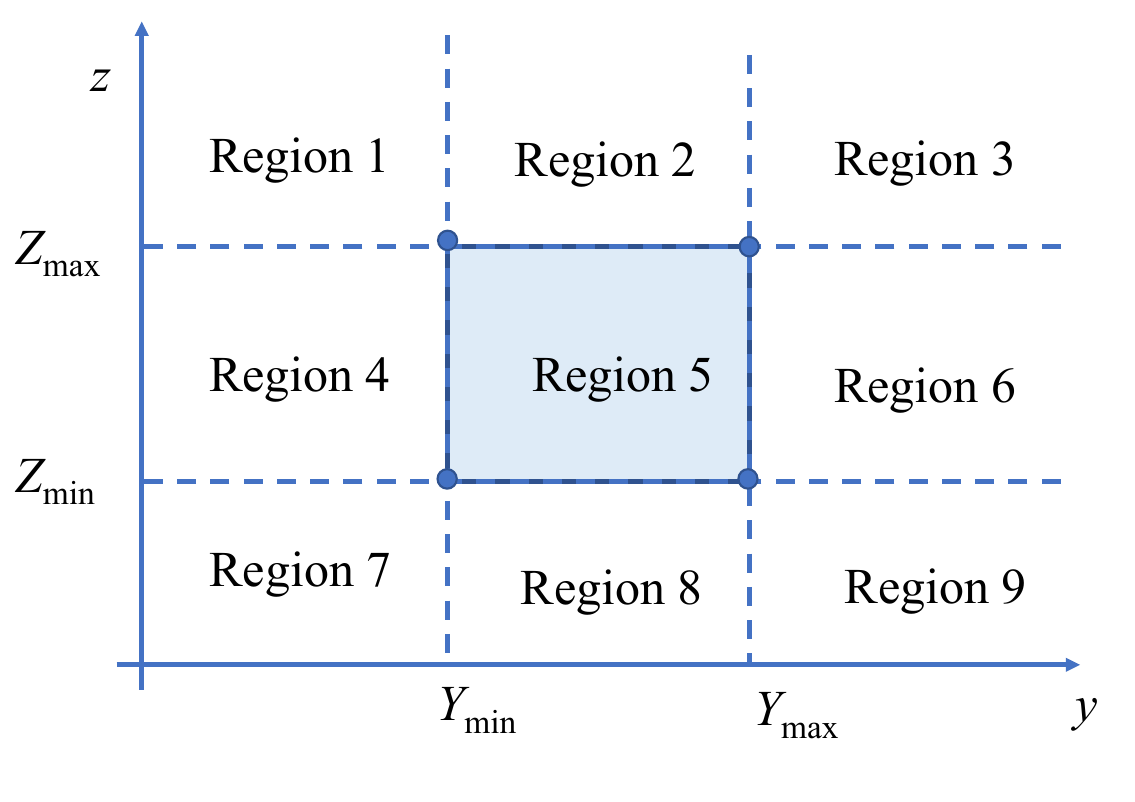}
    \caption{An illustration of the optimal region of the antenna location for FAR port B.}
    \label{fig2}
\end{figure}

Observing that maximizing the channel gain for each user mainly lies in minimizing the distance from port B to the BS as shown in \eqref{sys2eq1}.
As a result, the antenna location optimization problem \eqref{alo3max31}  for port B can be equivalent to 
\begin{subequations}\label{alo3max5} 
	\begin{align}
		\mathop{\min}_{y_2, z_2} \;&  
		 \sqrt{(X-s_1)^2+(y_2-s_2)^2+(z_2-H)^2}
      \\
		\textrm{s.t.} \quad\: 
		 & Y_{\min}\leq y_2\leq Y_{\max}, 
		\\&  Z_{\min}\leq z_2\leq Z_{\max}. 
	\end{align}
\end{subequations}
Due to the convex objective function and linear constraints, 
problem \eqref{alo3max5} is convex and the optimal solution can be effectively obtained. 
Through calculating the first-order derivative and constraints (\ref{alo3max5}b)-(\ref{alo3max5}c),
the optimal solution of problem \eqref{alo3max5} can be obtained as 
\begin{align}\label{alo31eq61}
   y_2^*=s_2|_{Y_{\min}}^{Y_{\max}},
\end{align}
and 
\begin{align}\label{alo31eq62}
   z_2^*=H|_{Z_{\min}}^{Z_{\max}},
\end{align}
where $a|_b^c=\min\{\max\{a,b\},c\}$.
Based on \eqref{alo31eq61} and \eqref{alo31eq62}, the optimal region of  the antenna location for FAR port B can be shown in Fig.~\ref{fig2}.

\subsubsection{Antenna Location Optimization for Port A}
Due to the complicated objective function (\ref{alo3max32}a) and constraint (\ref{alo3max32}b), it is hard to solve problem \eqref{alo3max32}.
To handle this issue, we introduce a slack variable $q_n$ and problem \eqref{alo3max32} can be made equivalent to
\begin{subequations}\label{alo3max7} 
	\begin{align}
		\mathop{\max}_{ y_1, z_1, \boldsymbol q} \;&  
		 \left(B-\sum_{n \in \mathcal N, n \neq k}\frac{R_n}{q_n}\right)
 q_k
   \\
		\textrm{s.t.} \quad\:
		&  B-\sum_{n \in \mathcal N, n \neq k} \frac{R_n}{q_n}\geq 0 , \\
  &q_n\leq {\log_2\left(1+ \frac{p_n h_{n}( y_1, z_1,y_2, z_2)}{\sigma^2}\right)}, \quad \forall n\\
  &q_n\geq 0,\quad \forall n
		\\& Y_{\min}\leq y_1\leq Y_{\max}, 
		\\&  Z_{\min}\leq z_1\leq Z_{\max},
	\end{align}
\end{subequations}
where $\boldsymbol q=[q_{1},\cdots,q_{N}]^T$.
Although the form of the objective function (\ref{alo3max7}a) is in a simple form, (\ref{alo3max7}a) is still non-convex. 
Recalling that the Hessian matrix of function $\frac{q_k^2}{q_n}$
\begin{align}\label{alo3eq8}
\left [ \begin{matrix} 
    \frac{\partial^2 \frac{q_k^2}{q_n}}{\partial q_k^2} &  \frac{\partial^2 \frac{q_k^2}{q_n}}{\partial q_k\partial q_n}  \\
  \frac{\partial^2 \frac{q_k^2}{q_n}}{\partial q_k\partial q_n}    &  \frac{\partial^2 \frac{q_k^2}{q_n}}{\partial q_n^2} 
\end{matrix} \right ]
= \frac{2}{q_n^3} 
\left [ \begin{matrix} 
    q_n  \\
  -q_k 
\end{matrix} \right ]
\left [ \begin{matrix} 
    q_n  &
  -q_k 
\end{matrix} \right ]
\end{align}
is positive semi-definite, 
function $\frac{q_k^2}{q_n}$ is convex.
Through replacing $q_k$ with $q_k^2$, problem \eqref{alo3max7}  is further equivalent to
\begin{subequations}\label{alo31max8} 
	\begin{align}
		\mathop{\max}_{ y_1, z_1, \boldsymbol q} \;&  
		 \left(B-\sum_{n \in \mathcal N, n \neq k}\frac{R_n}{q_n}\right)
 q_k^2
   \\
		\textrm{s.t.} \quad\:
		&  B-\sum_{n \in \mathcal N, n \neq k} \frac{R_n}{q_n}\geq 0 , \\
  &q_n\leq {\log_2\left(1+ \frac{p_n h_{n}( y_1, z_1,y_2, z_2)}{\sigma^2}\right)}, \quad \forall n\neq k\\
  &q_k^2\leq {\log_2\left(1+ \frac{p_k h_{k}( y_1, z_1,y_2, z_2)}{\sigma^2}\right)},  \\
  &q_n\geq 0,\quad \forall n
		\\& Y_{\min}\leq y_1\leq Y_{\max}, 
		\\&  Z_{\min}\leq z_1\leq Z_{\max}.
	\end{align}
\end{subequations}
Consequently, the objective function (\ref{alo31max8}a) has the form of the difference of two convex functions.
Taking the first-order derivative, objective function (\ref{alo31max8}a) can be approximated by
\begin{align}\label{alo31eq63}
B (q_k^{(t)})^2+2B q_k^{(t)}(q_k-q_k^{(t)})-\sum_{n \in \mathcal N, n \neq k}\frac{R_n}{q_n}
 q_k^2, \end{align}
where the superscript $(t)$ denotes the value of the variable in the $t$-th iteration. 
Now, it remains to handle the difficulty of constraints 
(\ref{alo31max8}c) and (\ref{alo31max8}d). 
Introducing a slack variable $u_n$, constraints 
(\ref{alo31max8}c) and (\ref{alo31max8}d) can be represented by 
\begin{align}\label{alo31eq71}
q_n\leq  \log_2\left(1+  u_n\right), \quad \forall n\neq k,
 \end{align}
 \begin{align}\label{alo31eq712}
u_n\leq \frac{p_{n} h_{n}( y_1, z_1,y_2, z_2)}{\sigma^2}, \quad \forall n,
 \end{align}
 \begin{align}\label{alo31eq72}
q_k^2\leq  \log_2\left(1+  u_k\right).
\end{align}
Substituting  \eqref{sys2eq1} into \eqref{alo31eq712} yields 
 \begin{align}\label{alo31eq8}
 \nonumber
  &\sigma^2({p_n\rho_0)}^{-1} \left( d_{n1}( y_1, z_1)+\frac{d_2( y_1, z_1, y_2, z_2)}{A}+d_3(y_2, z_2)\right)^{\alpha}\\
  &\leq \frac 1 {u_n}.
\end{align}
Through replacing the right hand side with its first-order derivative, constraint \eqref{alo31eq8} can be transformed to 
 \begin{align}\label{alo31eq9}
  &\sigma^2({p_n\rho_0)}^{-1} \left( d_{n1}( y_1, z_1)+\frac{d_2( y_1, z_1, y_2, z_2)}{A}+d_3(y_2, z_2)\right)^{\alpha}
  \nonumber\\
  &\leq \frac 1 {u_n^{(t)}}-\frac 1 {(u_n^{(t)})^2} (u_n-u_n^{(t)}).
\end{align}
Denote  $\boldsymbol u=[u_{1},\cdots,u_{N}]^T$.
As a result, with approximations \eqref{alo31eq63}
and \eqref{alo31eq9}, problem \eqref{alo31max8} can be transformed to 
\begin{subequations}\label{alo31max9} 
	\begin{align}
		\mathop{\max}_{ y_1, z_1, \boldsymbol q, \boldsymbol u } \;&  
B (q_k^{(t)})^2+2B q_k^{(t)}(q_k-q_k^{(t)})-\sum_{n \in \mathcal N, n \neq k}\frac{R_n}{q_n}
 q_k^2
   \\
		\textrm{s.t.} \quad\:
		&  B-\sum_{n \in \mathcal N, n \neq k} \frac{R_n}{q_n}\geq 0 , \\
  &q_n\leq {\log_2\left(1+ u_n\right)}, \quad \forall n\neq k\\
  &q_k^2\leq {\log_2\left(1+ u_k\right)},  \\
  &\sigma^2({p_n\rho_0)}^{-1} \left( d_{n1}( y_1, z_1)+\frac{d_2( y_1, z_1, y_2, z_2)}{A} \right.
  \nonumber\\
  &+d_3(y_2, z_2)\Bigg)^{\alpha}\leq \frac 1 {u_n^{(t)}}-\frac 1 {(u_n^{(t)})^2} (u_n-u_n^{(t)}), \quad \forall n,\\
  &q_n\geq 0,\quad \forall n
		\\& Y_{\min}\leq y_1\leq Y_{\max}, 
		\\&  Z_{\min}\leq z_1\leq Z_{\max},
	\end{align}
\end{subequations}
which is convex and can be solved using existing convex optimization tools such as CVX. 

\subsection{Algorithm Analysis}
As a result, the overall algorithm to solve problem \eqref{sys2max0} is shown in Algorithm 1.
Based on Algorithm 1, the main complexity lies in solving the convex problem  \eqref{alo31max9}, which involves the complexity of $\mathcal O(N^3)$  with the interior-point method for solving the convex problem.
As a result, the overall complexity of solving  problem \eqref{sys2max0} is $\mathcal O(IKN^3)$, where $I$ is the number of iterations for the successive convex approximation method.

 \begin{algorithm}[t]
 	\caption{Alternating Algorithm for Problem \eqref{sys2max0}}
 	\begin{algorithmic}[1] 
 		\STATE Obtain the optimal antenna location for FAR port B according to equations \eqref{alo31eq61} and \eqref{alo31eq62}.  
 		\FOR {$k=1, \cdots, N$}
 		\STATE Solve problem \eqref{alo31max8} through solving a set of convex subproblems \eqref{alo31max9} using the successive convex approximation method.
    \ENDFOR
    \STATE Choose the user $k$ with the highest sum rate and output the optimal bandwidth based on \eqref{alo31eq2} and \eqref{alo31eq3}. 
 	\end{algorithmic}
 \end{algorithm}

\addtolength{\topmargin}{0.02in}
\section{Simulation Results}\label{section4}
In this section, simulation results are conducted to verify the effectiveness of the proposed algorithm.
There are five users unified distributed in the square area with size 300m$\times$300m.
The BS is located at [350m, 30m, 30m].
The FAR with two ports is located between users and the BS.
We set the width of the wall configured with FAR as 20m.
The feasible region of both ports A and B in the FAR is [0m, 20m]$\times$[0m, 20m].
The bandwidth of the system is 10 MHz.
The transmit power of each user is set as the same, i.e., $p_1=\cdots=p_N$.
To show the effectiveness of the proposed scheme, we consider the following two baselines: the fixed fluid antenna location scheme, where the location of the antenna is in the center of the considered area, and the equal bandwidth scheme, where the bandwidth is equally assigned with all users. 

\begin{figure}[t]
    \centering
    \includegraphics[width=0.85\linewidth]{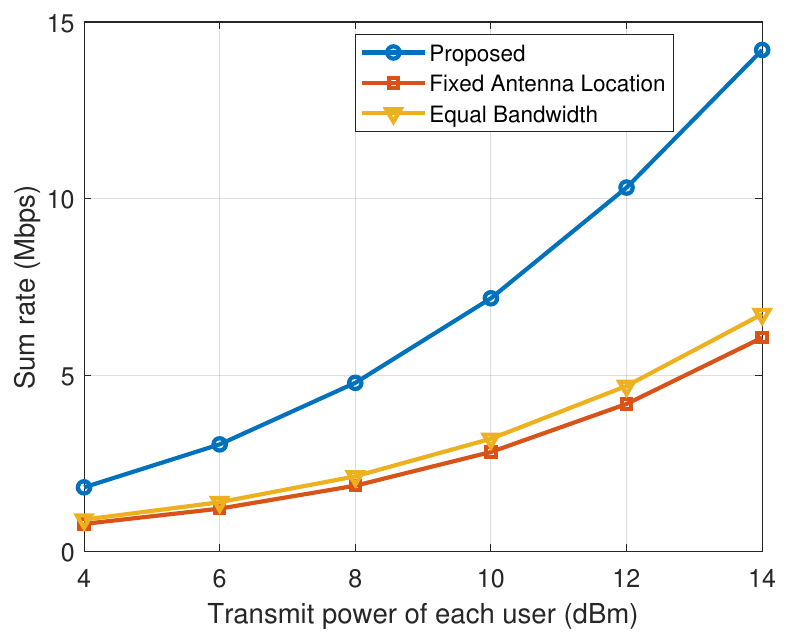}
    \caption{Sum rate versus the transmit power of each user.}
    \label{simfig1}
\end{figure}

Fig.~\ref{simfig1} illustrates the sum rate versus the transmit power of each user. 
It can be observed that the sum rate of all schemes increases with the transmit power.
The proposed algorithm achieves the best sum rate in particular for high transmit power region. 
The reason is that compared to the two baselines, the proposed algorithm jointly optimizes the bandwidth and the fluid antenna location, which shows the effectiveness of the proposed algorithm.
It is also found that the performance of the fixed antenna location scheme achieves the worst performance, which indicates the importance of the antenna location optimization.
Compared to the  fixed antenna location scheme, the sum rate can be increased to 125\% through the proposed scheme. 

\section{Conclusion}\label{section5}
In this paper, we  investigated the FAR assisted uplink wireless communication system with multiple users.
An optimization problem with jointly optimizing antenna location and bandwidth allocation was formulated.
The original problem was first transformed to an equivalent problem with only  antenna location variable through observing the optimal structure of the bandwidth allocation. Then, an alternating algorithm with low complexity was proposed. 
Simulation results verify the theoretical findings of the proposed algorithm. 
Future directions about FAR assisted communication system include the antenna location optimization for multiple FARs, the BS with multiple antennas, and the near-field considerations. 


\bibliographystyle{IEEEtran}
\bibliography{FAR}
\end{document}